\documentclass[a4paper,12pt]{article}

\usepackage{epsfig,graphicx,xcolor,soul,amsbsy,amssymb,latexsym,amsfonts,amsmath,setspace}
\usepackage{pstricks}
\usepackage{color}

\usepackage{eurosym}
\usepackage[a4paper]{geometry}
\usepackage{graphicx}
\usepackage{subfigure}
\usepackage{tikz}
\usepackage{xcolor}
\usepackage{amsmath,amssymb,amsfonts,pstricks,setspace}

\numberwithin{equation}{section}

\newcommand{\bea}{\begin{eqnarray}\displaystyle}
\newcommand{\eea}{\end{eqnarray}}

\newcommand{\figref}[1]{Fig.~\protect\ref{#1}}

\title{\begin{flushright}{\vspace{-2.5cm}\small }\end{flushright}\vspace{0.8cm}
\bf{Elliptic CY3folds and Non-Perturbative Modular Transformation}\\[15pt]}
\author{\large \textsc{Amer Iqbal\footnote{\tt amer@alum.mit.edu},~ Khurram Shabbir\footnote{\tt khurramsms@gmail.com}}}
\date{}

\begin{document}

\maketitle

\begin{center}
\renewcommand{\thefootnote}{\fnsymbol{footnote}}\vspace{-0.5cm}
${}^{\footnotemark[1]}$ Abdus Salam School of Mathematical Sciences \\ Government College University, Lahore, Pakistan\\[0.4cm]
${}^{\footnotemark[2]}$ Department of Mathematics\\Government College University, Lahore, Pakistan\\[1cm]
\end{center}

\begin{abstract}
\noindent
We study the refined topological string partition function of a class of toric elliptically fibered Calabi-Yau threefolds. These Calabi-Yau threefolds give rise to five dimensional quiver gauge theories and are dual to configurations of M5-M2-branes. We determine the Gopakumar-Vafa invariants for these threefolds and show that the genus $g$ free energy is given by the weight $2g$ Eisenstein series. We also show that although the free energy at all genera are modular invariant the full partition function satisfies the non-perturbative modular transformation property discussed by Lockhart and Vafa in arXiv:1210.5909 and therefore the modularity of free energy is up to non-perturbative corrections.

\end{abstract}

${}$\\[500pt]

\onehalfspacing
\vskip1cm


\section{Introduction}
In this paper we study the refined topological string partition function of a class of toric elliptically fibered Calabi-Yau threefolds which are dual to a set of parallel M5-branes with a transverse direction compactified to a circle. The size of the circle is related to the K\"ahler parameter of the elliptic curve class on the Calabi-Yau threefolds. We determine the Gopakumar-Vafa invariants for all curve classes and also show that for these Calabi-Yau threefolds the genus $g$ free energy takes a particularly simple for and is given by the Eisenstein series. We also study the modular properties of the refined partition function and  show that it satisfies the non-perturbative modular transformation discovered in \cite{Lockhart:2012vp}. As was shown in \cite{Lockhart:2012vp} for the case of $N=1$, $N$ being the number of M5-branes, the partition function is not modular invariant but satisfies a more involved transformation which maps the $g_{s}$ to $\frac{1}{g_{s}}$ where $g_{s}$ is the topological string coupling constant hence the name non-perturbative modular transformation.

This note is organized as follows. In section 2 we discuss the Calabi-Yau geometry and the dual brane configuration. In section 3 we use the refined topological vertex formalism to express the partition function as a trace of an operator acting on the fermionic Fock space. In section 4 we determine the Gopakumar-Vafa invariants for all curve classes and the genus $g$ fee energy. In section 5 we express the partition function in terms of double elliptic gamma functions  and show that it satisfies the non-perturbative modular transformation property. 

\section{Elliptic CY3fold and dual brane configuration}

The class elliptic Calabi-Yau threefolds we are interested in are dual to certain brane configurations which arose in the study of M-strings \cite{mstrings}. These Calabi-Yau threefolds are birationally equivalent $\widehat{A}_{N-1}\times_{f}\mathbb{C}$ where $\widehat{A}_{N-1}$ is affine $A_{N-1}$ space blown up at $N$ points and it is fibered over $\mathbb{C}$ to obtain the Calabi-Yau threefold. We will denote these by $X_{N}$.

The affine $A_{N-1}$ space has $N$ $\mathbb{P}^1$'s corresponding to the simple roots of affine $SU(N)$. We denote these curve classes by $C_{a}$ with $a=1,2,\cdots, N$. The blow-up introduces $N$ new curve classes which we denote by $M_{a}$ with $a=1,2,\cdots,N$. The class of the elliptic curve $E$ is given by
\bea\label{EC}
E=C_{1}+C_{2}+\cdots+C_{N}\,.
\eea
The complexified K\"ahler parameters associated with these curves are given by the K\"ahler form $\omega$:
\bea
t_{a}=\int_{C_{a}}\,\omega\,,\,\,\,\rho=\int_{E}\,\omega\,,\,\,m=\int_{M_{a}}\,\omega\,,\,\,\,\,a=1,2,\cdots,N\,.
\eea
Because of Eq.(\ref{EC}) $\rho$ is given by,
\bea
\rho=t_{1}+t_{2}+\cdots+t_{N}\,.
\eea

\begin{figure}[h]
\centering
  \includegraphics[width=5in]{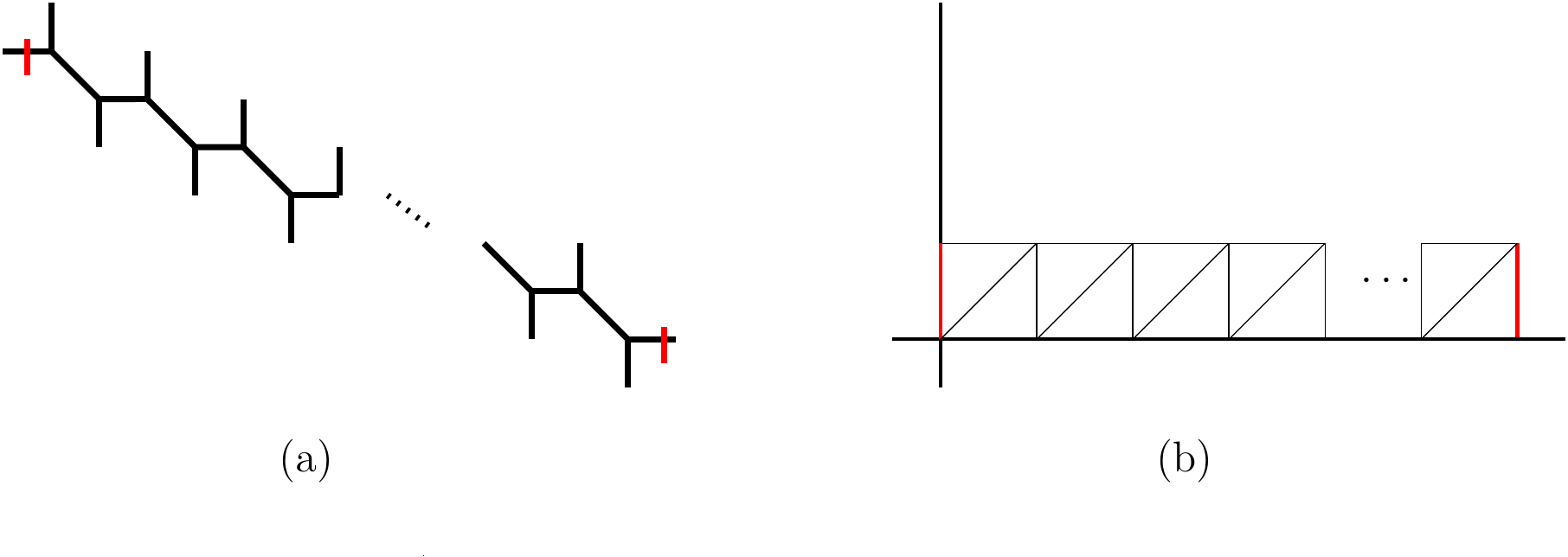}
  \caption{(a) The $(p,q)$ brane configuration dual to $X_{N}$. The web lives on $\mathbb{R}\times S^1$ and the horizontal lines are glued together. (b) The Newton polygon of $X_{N}$. }\label{figure1}
\end{figure}

The $(p,q)$ 5-brane web dual to $X_{N}$ and the Newton polygon are given by \figref{figure1}(a) and \figref{figure1}(b) respectively.

The Calabi-Yau threefolds $X_{N}$ are dual to certain M5-brane configurations. Consider $N$ M5-brane with worldvolume coordinates $X^{A},A=0,1,2,3,4,5$. The transverse space is $\mathbb{R}^{5}$ with coordinates $X^{I},I=6,7,8,9,10$. We consider the $X^6$ direction and separate the $N$ M5-branes along this direction and then compactify this direction to a circle so that M5-branes are sprinkled on the $X^6$ circle. The transverse space to the M5-branes is $\mathbb{R}^{4}_{\perp}\times S^1$ and one can introduce a deformation by fibering this transverse space over the $X^6$ circle so that as one goes around the $X^6$ circle there is an $SO(4)$ action on $\mathbb{R}^4_{\perp}$ given by \cite{mstrings},
\bea\label{mass}
(w_{1},w_{2})\mapsto (e^{2\pi i m}w_{1},e^{-2\pi im}w_{2})\,.
\eea
The deformation $m$ breaks the maximal supersymmetry of the brane configuration. In the type IIB picture the the theory living on the $(p,q)$ brane web is five dimensional quiver gauge theory with eight supercharges. The partition function of this gauge theory can be calculated using the Nekrasov's instanton calculus or by applying the refined topological vertex formalism to the $(p,q)$ brane web of \figref{figure1}(a).

\section{The partition function from topological vertex}

The partition function of the $(p,q)$ brane web shown in \figref{figure1} can be calculated by applying refined topological vertex formalism to the web. The refined vertex requires a choice of preferred direction \cite{Iqbal:2007ii} which determines the form of the partition function as a sum over Young diagrams but the partition function itself is invariant under the change in the preferred direction. In this case the simplest choice for the preferred direction is vertical and, as we will see below, in this case the sum can be carried out exactly and partition function becomes an infinite product. We denote the topological string partition function of this Calabi-Yau threefold by $Z_{N}$ and it is given by \cite{randompaper}
\bea\label{PF1}
Z_{N}:=\sum_{\vec{\lambda}}\prod_{a=0}^{N-1}\Big[(-Q_{a})^{|\lambda_{(a)}|}\,\Big(\sum_{\mu}(-Q_{m})^{|\mu|}\,C_{\lambda^{t}_{(a)}\,\mu\emptyset}(t, q)\,C_{\lambda_{(a+1)}\mu^{t}\emptyset}(q, t)\Big)\Big]\,,
\eea
where
\bea
C_{\lambda\mu\emptyset}(t,q)&=&\Big(\frac{q}{t}\Big)^{\frac{|\lambda|-|\mu|}{2}}\,\sum_{\eta}\Big(\frac{q}{t}\Big)^{\frac{|\eta|}{2}}\,s_{\lambda^{t}/\eta}(t^{-\rho})\,s_{\mu/\eta}(q^{-\rho})
\eea
is the refined topological vertex. The length of the slanted lines in \figref{figure1} are all equal to $m$ and we have defined $Q_{m}=e^{-m}$, similarly the length of the horizontal lines is $T_{a}$ and we have defined $Q_{a}=e^{-T_{a}}$ such that $t_{a}=T_{a}+m$ is the distance between the two vertical lines. $\lambda$, $\mu$ and $\eta$ are Young diagrams and $s_{\lambda/\mu}(x_{1},x_{2},\cdots)$ is the skew-Schur polynomial in variables $x_{1},x_{2},\cdots$. $t^{-\rho}$ denotes the following set of variables: $t^{-\rho}=\{t^{1/2},t^{3/2},\cdots\}$. 

Using repeatedly the  identity
\bea\label{id1}
\sum_{\lambda}s_{\lambda^{t}/\eta}(x)s_{\lambda/\sigma}(y)&=&\prod_{i,j}(1+x_{i}y_{j})\, \sum_{\tau}s_{\sigma^{t}/\tau}(x)\,s_{\eta^{t}/\tau^{t}}(y)
\eea
and the following properties of the skew-Schur functions,
\bea\label{id2}
s_{\lambda^t/\sigma^t}(q^{-\rho})=s_{\lambda/\sigma}(-q^{\rho})\,,\,\,\,\sum_{\eta}s_{\lambda/\eta}(x)\,s_{\eta/\sigma}(y)=s_{\lambda/\sigma}(x,y),
\eea
the partition function in Eq.(\ref{PF1}) can be written as,
\bea
Z_{N}=\Pi(Q_{m})^{N}\sum_{\vec{\lambda}, \vec{\tau}} Q_{\rho}^{|\lambda_{(0)}|}\,\prod_{a=0}^{N-1}s_{\lambda_{(a)}/\tau_{(a+1)}}({\bf x}_{a+1})\,s_{\lambda_{(a+1)}/\tau_{(a+1)}}({\bf y}_{a+1})\,,\label{PFPF}
\eea
where:
\bea \nonumber 
Q_{\rho} &=& \prod\limits_{a=0}^{N-1} (Q_a Q_m)\quad ,\quad Q_{1,a+1}=(Q_{1}Q_{2}\cdots Q_{a})Q_{m}^{a}\\\nonumber
{\bf x}_{a+1}&=& Q_m^{-1/2} Q_{1,a+1}\,\{Q_{m}\sqrt{\frac{q}{t}}t^{\rho},t^{-\rho}\}\,,
{\bf y}_{a+1}= Q_m^{-1/2} Q_{1,a+1}^{-1}\,\{Q_{m}\sqrt{\frac{t}{q}}q^{-\rho},q^{\rho}\}
\,.
\eea
and 
\bea\label{pidef}
\Pi(x)=\prod_{i,j=1}^{\infty}\Big(1-x\,q^{-\rho_{i}}\,t^{-\rho_{j}}\Big)\,.
\eea
In the above equations $-log(Q_{\rho})$ is the K\"ahler parameter associated to the elliptic curve which is dual to the circle transverse to the M5-branes. The form of the partition function given in Eq.(\ref{PFPF}) was obtained in \cite{randompaper} where it was related to the periodic Schur process of period $N$.

\subsection{Product representation of partition function using free fermions}

In this section we express the partition function given in Eq.(\ref{PFPF}) as an infinite product. To do this, following \cite{okounkov} we introduce the free fermion Fock space spanned by creation and annihilation operators $(\psi_a,\psi_a^*)$ satisfying the relations
\bea
&\{\psi_a,\psi_b\}=\{\psi_a^*,\psi_b^*\}=0\,,\,\,\,\,\{\psi_a,\psi_b^*\}=\delta_{ab}\,,\,a,b\in \mathbb{Z}+\tfrac{1}{2}\,
\eea
With these we can construct the operators
\bea\label{ops}
\Gamma_{\pm}(z)=\mbox{exp}\left(\sum_{n=1}^\infty\frac{z^n J_n}{n}\right)\,,\,\,\,\,
\mbox{where}\,\,\,\,\,
J_n=\sum_{k\in\mathbb{Z}+\frac{1}{2}}\psi_{k+n}\psi^*_{k}\,,\,\,\,\,\,\,\,\,n=\pm1,\pm2,\ldots\,.
\eea
These operators satisfy the following commutation relation,
\bea
\Gamma_+(z)\Gamma_-(w)=(1-zw)\Gamma_-(w)\Gamma_+(z)\,.
\eea
The fermionic Fock space is spanned by states which are in one-to-one correspondence with partitions. For a partition $\lambda=(\lambda_{1},\lambda_{2},\cdots)$ we have the state,
\bea
|\lambda\rangle=\psi_{\lambda_{1}-\frac{1}{2}}\psi_{\lambda_{2}-\frac{3}{2}}\psi_{\lambda_{3}-\frac{5}{2}}\cdots\,|0\rangle\,.
\eea
The ground state $|0\rangle$ is annihilated by $\psi_{k}$ for $k<0$. The operators defined in Eq.(\ref{ops}) are useful when working with symmetric polynomials due to the following,
\bea\label{schurcreation}
\prod_{i}\Gamma_{+}(x_{i})|\lambda\rangle&=&\sum_{\mu}s_{\mu/\lambda}(x_{1},x_{2},\cdots)|\mu\rangle\,\\\nonumber
\prod_{i}\Gamma_{-}(x_{i})|\lambda\rangle&=&\sum_{\mu}s_{\lambda/\mu}(x_{1},x_{2},\cdots)|\mu\rangle\,.
\eea
To simplify the equations we will use the following notation: $\prod_{i}\Gamma_{\pm}(x_{i})=\Gamma_{\pm}({\bf x})$. Using Eq.(\ref{PFPF}) and Eq.(\ref{schurcreation}) we can write $\widehat{Z}_{N}:=Z_{N}/\Pi(Q_{m})^N$ as trace, over the free fermionic Fock space, of infinite product of operators:
\begin{align}\label{eq1}
\widehat{Z}_{N}=\mbox{Tr}\Big((Q_{1}Q_{m})^{L_{0}}{\cal O}(Q_{2}Q_{m})^{L_{0}}{\cal O}\cdots (Q_{N}Q_{m})^{L_{0}}{\cal O}\Big)\,,
\end{align}
with the operator insertions $\mathcal{O}$ being built from $\Gamma_\pm$
\begin{align}
{\cal O}=\prod_{i=1}^{\infty}\Big(\Gamma_{+}(Q_{m}^{-1}\,t^{-i+\frac{1}{2}})\Gamma_{+}(\sqrt{\frac{q}{t}}t^{i-\frac{1}{2}}\Big)\,\prod_{j=1}^{\infty}\Big(\Gamma_{-}(q^{j-\frac{1}{2}})\Gamma_{-}(Q_{m}\sqrt{\frac{t}{q}}q^{-j+\frac{1}{2}})\Big)\,.
\end{align}
In the above equations $L_{0}$ is such that $L_{0}|\lambda\rangle=|\lambda|\,|\lambda\rangle$.
Using the commutation relations of $\Gamma_{\pm}({\bf x})$, we can write Eq.(\ref{eq1}) as,
\begin{align}
\widehat{Z}_{N}=\mbox{Tr}\Big(Q_{\rho}^{L_{0}}\prod_{a=1}^{k}\Gamma_{+}(Q_{1}Q_{2}\cdots Q_{a}Q_{m}^{a}{\bf x})\Gamma_{-}(Q_{1}^{-1}Q_{2}^{-1}\cdots Q_{a}^{-1}Q_{m}^{-a}{\bf y})\Big)\,,
\end{align}
where
\begin{align}
&{\bf x}=\{Q_{m}^{-1}t^{-i+\frac{1}{2}},\sqrt{\frac{q}{t}}t^{i-\frac{1}{2}}\,|\,i=1,2,\cdots\}\,,&&{\bf y}=\{q^{i-\frac{1}{2}},Q_{m}\sqrt{\frac{t}{q}}q^{-i+\frac{1}{2}}\,|\,i=1,2,\cdots\}\,.
\end{align}
Using commutation relation of $\Gamma_{\pm}(x)$ repeatedly we get,
\begin{align}\label{eq2}
\widehat{Z}_{N}=\Big(\prod_{1\leq a<b\leq k}F_{ab}\Big)\mbox{Tr}\Big(Q_{\rho}^{L_{0}}\Gamma_{+}({\bf X})\Gamma_{-}({\bf Y})\Big)\,,
\end{align}
where we have denoted
\begin{align}
{\bf X}&=\{Q_{1}Q_{m}{\bf x},Q_{1}Q_{2}Q_{m}^{2}{\bf x},\cdots,
Q_{1}Q_{2}\cdots Q_{N}Q_{m}^{N}{\bf x}\}\\
{\bf Y}&=\{Q_{1}^{-1}Q_{m}^{-1}{\bf y},Q_{1}^{-1}Q_{2}^{-1}Q_{m}^{-2}{\bf y},\cdots,
Q_{1}^{-1}Q_{2}^{-1}\cdots Q_{N}^{-1}Q_{m}^{-N}{\bf y}\}\\
F_{ab}&=\prod_{i,j=1}^{\infty}\frac{(1-Q_{ab}Q_{m}^{-1}t^{i-\frac{1}{2}}q^{j-\frac{1}{2}})(1-Q_{ab}Q_{m}t^{i-\frac{1}{2}}q^{j-\frac{1}{2}})}
{(1-Q_{ab}t^{i}q^{j-1})(1-Q_{ab}t^{i-1}q^{j})}\,,\\
Q_{ab}&=Q_{a+1}\cdots Q_{b}Q_{m}^{b-a}\,.
\end{align}
The trace appearing in Eq.(\ref{eq2}) can be written in the form of a product \cite{macdonald},
\begin{align}
\mbox{Tr}\Big(Q_{\rho}^{L_{0}}\Gamma_{+}({\bf x})\Gamma_{-}({\bf y})\Big)=\prod_{n=1}^{\infty}(1-Q_{\rho}^{n})^{-1}\prod_{i,j}(1-Q_{\rho}^{n}x_{i}y_{j})^{-1}\,.
\end{align}
such that the partition function is given by
\bea\label{kbrane}
\widehat{Z}_{N}&=(\prod_{n=1}^{\infty}(1-Q_{\rho}^{n})^{-1})(\prod_{1\leq a<b\leq N}F_{ab}\Big)(\prod_{a,b=1}^{N}H_{ab}\Big)\,,
\eea
where
\begin{align}\label{eq3}
H_{ab}&=\prod_{n,i,j=1}^{\infty}\frac{(1-Q_{\rho}^{n}\widetilde{Q}_{ab}Q_{m}^{-1}\,t^{i-\frac{1}{2}}q^{j-\frac{1}{2}})(1-Q_{\rho}^{n}\widetilde{Q}_{ab}Q_{m}\,t^{i-\frac{1}{2}}q^{j-\frac{1}{2}})}
{(1-Q_{\rho}^{n}\widetilde{Q}_{ab}t^{i}q^{j-1})(1-Q_{\rho}^{n}\widetilde{Q}_{ab}t^{i-1}q^{j})}\,,\\\nonumber
\widetilde{Q}_{ab}&=Q_{1}Q_{2}\cdots Q_{a}Q_{1}^{-1}\cdots Q_{b}^{-1}Q_{m}^{a-b}
\end{align}

\section{Free energy and Gopakumar-Vafa invariants}
Using the refined topological string partition function we can calculate the free energy and the Gopakumar-Vafa invariants \cite{Gopakumar:1998ii,Gopakumar:1998jq} for the different curve classes. The free energy is given by,
\bea
F_{N}(\rho,t_{1},\cdots,t_{N-1},m)&=&\mbox{ln}\,Z_{N}\\\nonumber
&=&N\,\mbox{ln}\,\Pi(Q_{m})+\mbox{ln}\widehat{Z}_{N}\,,
\eea
where $\Pi(Q_m)$ is given by Eq.(\ref{pidef}) and $\widehat{Z}_{N}$ is given by Eq.(\ref{kbrane}). After some simplification $F_{N}$ can be written as,
\bea\label{freeenergy}
F_{N}(\rho,t_{1},\cdots,t_{N-1},m)&=&-N\,\sum_{k=1}^{\infty}\frac{Q_{m}^{k}}{k}\,\Big[\frac{1}{(q^{\frac{k}{2}}-q^{-\frac{k}{2}})(t^{\frac{k}{2}}-t^{-\frac{k}{2}})}\Big]+\sum_{n=1}^{\infty}\sum_{k=1}^{\infty}\frac{Q_{\rho}^{n\,k}}{k}\\\nonumber &+&\sum_{1\leq a<b\leq N}\sum_{k=1}^{\infty}\frac{Q_{ab}^{k}}{k}\,\Big[\frac{(t/q)^{\frac{k}{2}}+(q/t)^{\frac{k}{2}}-(Q_{m}^{k}+Q_{m}^{-k})\,}{(q^{\frac{k}{2}}-q^{-\frac{k}{2}})(t^{\frac{k}{2}}-t^{-\frac{k}{2}})}\Big]\\\nonumber
&+&\sum_{n=1}^{\infty}\sum_{a,b=1}^{N}\sum_{k=1}^{\infty}\frac{Q_{\rho}^{n\,k}\widetilde{Q}_{ab}^{k}}{k}\Big[\frac{(t/q)^{\frac{k}{2}}+(q/t)^{\frac{k}{2}}-(Q_{m}^{k}+Q_{m}^{-k})\,}{(q^{\frac{k}{2}}-q^{-\frac{k}{2}})(t^{\frac{k}{2}}-t^{-\frac{k}{2}})}\Big]\,.
\eea
From Eq.(\ref{freeenergy}) we can see that the $SU(2)_{L}\times SU(2)_{R}$ spin content of various curve classes is given by:
\begin{center}
\begin{tabular}{||p {10cm}|p {4cm}||}\hline
Curve & $\sum_{j_{L},j_{R}}N_{C}^{j_{L},j_{R}}(j_{L},j_{R})$\\\hline
$n\,E+C_{ab}\,,\,\,n\geq 0,\,1\leq a<b\leq N,\,c=a-1,b+1$ & $(0,\frac{1}{2})$\\\hline
$n\,E+C_{ab}+M_{c}\,,\,\,n\geq 0,\,1\leq a<b\leq N,\,c=a-1,b+1$ & $(0,0)$\\\hline
$n\,E+C_{ab}-M_{c}\,,\,\,n\geq 0,\,1\leq a<b\leq N,\,c=a+1,b-1$ & $(0,0)$\\\hline
$n\,E-C_{ab}\,,\,\,n\geq 1,\,1\leq a<b\leq N$ & $(0,\frac{1}{2})$\\\hline
$n\,E-C_{ab}+M_{c}\,,\,\,n\geq 1,\,1\leq a<b\leq N,\,c=a-1,b+1$ & $(0,0)$\\\hline
$n\,E-C_{ab}-M_{c}\,,\,\,n\geq 1,\,1\leq a<b\leq N,\,c=a+1,b-1$ & $(0,0)$\\\hline
$n\,E\,,\,\,n\geq 1$ & $(\frac{1}{2},0)\oplus (N-1)(0,\frac{1}{2})$\\\hline
\end{tabular}
\end{center}
Where 
\bea\nonumber
C_{ab}=C_{a}+C_{a+1}+\cdots+C_{b}\,.
\eea

From Eq.(\ref{freeenergy}) we can isolate the contribution of the elliptic curve $E$ to $F_{N}$, which we will denote by $F^{E}_{N}$, by considering only those terms which depend only on $Q_{\rho}$ and other K\"ahler parameter,
\bea
F^{E}_{N}(\rho,q,t)&=&\sum_{n,k=1}^{\infty}\frac{Q_{\rho}^{n\,k}}{k}\Big[1+N\Big(\frac{t^{k}+q^{k}}{(1-q^{k})(1-t^k)}\Big)\Big]\,.
\eea
We will restrict ourselves to the unrefined case so that $q=t$ and in terms of the topological string coupling constant $g_{s}$,
\bea
q=t=e^{i\,g_{s}}\,.
\eea
Since 
\bea
\frac{2q}{(1-q)^2}&=&\frac{2}{g_{s}^2}-\frac{1}{6}-2\sum_{g\geq 2}^{\infty}\,g_{s}^{2g-2}\,\frac{B_{2g}}{(2g)(2g-2)!}\,,
\eea
where $B_{k}$ are the Bernoulli numbers therefore
\bea
F^{E}_{N}(\rho,q)=\sum_{g=0}^{\infty}g_{s}^{2g-2}F^{E}_{N,g}(\rho)\,
\eea
with
\bea
F^{E}_{N, g\geq 2}&=&N\frac{B_{2g}B_{2g-2}}{2(2g)\,(2g-2)!}-2\,N\frac{B_{2g}}{(2g)\,(2g-2)!}\,\,\frac{B_{2g-2}}{4}\,E_{2g-2}(\rho)\,,\\\nonumber
F^{E}_{1}&=&-\frac{1}{12}\mbox{ln}\Big[\prod_{n=1}^{\infty}(1-Q_{\rho}^n)^{12-2N}\Big]\,,\,\,\frac{\partial^2\,F^{E}_{0}}{\partial \rho^2}=2N\,\sum_{n=1}^{\infty}n^2\,\mbox{ln}(1-Q_{\rho}^n)^{-1}
\eea
where $E_{2g}(\rho)$ is the Eisenstein series defined as:
\bea
E_{2g}(\rho)=1-\frac{4}{B_{2g}}\sum_{n\geq 1}n^{2g-1}\,\Big(\frac{Q_{\rho}^n}{1-Q_{\rho}^n}\Big)\,.
\eea

\section{Non-perturbative modular transformation}

For an elliptic Calabi-Yau threefold with modular parameter $\rho$ one naively would expect the partition function to be invariant under the modular transformation,
\bea
Z_{N}\Big(-\frac{1}{\tau},\frac{\epsilon_1}{\tau},\frac{\epsilon_2}{\tau}\Big)=Z_{N}\Big(\tau,\epsilon_1,\epsilon_2\Big)
\eea
where $q=e^{i\epsilon_{1}}$ and $t=e^{-i\epsilon_{2}}$. However, quite surprisingly in \cite{Lockhart:2012vp} it was shown that the refined topological partition function of $X_{1}$, which is dual to a single M5-brane wrapped on a circle, satisfies a non-perturbative modular transformation,
\bea\label{NP}
Z_{1}\Big(-\frac{1}{\tau},\frac{\epsilon_1}{\tau},\frac{\epsilon_2}{\tau}\Big)=\frac{Z_{1}(\tau,\epsilon_1,\epsilon_2)}{Z_{1}(\frac{\tau}{\epsilon_1},-\frac{1}{\epsilon_1},\frac{\epsilon_2}{\epsilon_1})Z_{1}(\frac{\tau}{\epsilon_2},\frac{\epsilon_1}{\epsilon_2},-\frac{1}{\epsilon_2})}\,.
\eea
Since $\epsilon_1,\epsilon_2$ are related to the topological string coupling constant $g_{s}$ by
\bea
\epsilon_1\epsilon_2=-g_{s}^2\,,
\eea
therefore Eq.(\ref{NP}) implies that
\bea
Z_{1}\Big(-\frac{1}{\tau},\frac{\epsilon_1}{\tau},\frac{\epsilon_2}{\tau}\Big)=Z_{1}(\tau,\epsilon_1,\epsilon_2)+O(e^{-\frac{1}{g_{s}}})\,,
\eea
i.e., modularity holds only up to non-perturbative corrections. For this reason the transformation in Eq.(\ref{NP}) was called non-perturbative modular transformation and it was argued that this gives the non-perturbative completion of the topological string partition function for the case of elliptic Calabi-Yau threefolds.

To show that $Z_{N}$ also satisfies equation similar to Eq.(\ref{NP}) we write it in terms of double elliptic gamma functions. Recall that the double elliptic Gamma function is defined as,
\begin{align}
G_{2}(x;\tau,\epsilon_{1},\epsilon_{2})=\prod_{k,i,j=1}^{\infty}(1-Q_{\tau}^{k-1}q^{i-1}t^{-j+1}x)
(1-Q_{\tau}^{k}q^{i}t^{-j}x^{-1})
\end{align}
and satisfies the following modular transformation:
\begin{align}\label{pp}
G_{2}(z;\rho,\epsilon_1,\epsilon_2)&=&G_{2}\Big(\frac{z}{\rho};-\frac{1}{\rho},\frac{\epsilon_1}{\rho},\frac{\epsilon_2}{\rho}\Big)
G_{2}\Big(\frac{z}{\epsilon_1};\frac{\rho}{\epsilon_1},-\frac{1}{\epsilon_1},\frac{\epsilon_2}{\epsilon_1}\Big)
G_{2}\Big(\frac{z}{\epsilon_2};\frac{\rho}{\epsilon_2},\frac{\epsilon_1}{\epsilon_2},-\frac{1}{\epsilon_2}\Big)\,\mbox{exp}\Big(\frac{i\pi}{12}B_{44}\Big)\,,
\end{align}
where $B_{4,4}$ is given by
\begin{align}
B_{4,4}(z;\rho,\epsilon_1,\epsilon_2)=\frac{d^{4}}{dx^4}\frac{x^4\,e^{z\,x}}{(e^{\rho\,x}-1)(e^{\epsilon_{1}\,x}-1)(e^{\epsilon_{2}\,x}-1)}|_{x=0}\,.
\end{align}

The triple infinite product in Eq.(\ref{eq3}) can be written in terms of double elliptic gamma functions so that the full partition function $Z_N$ is given by,
\begin{align}\nonumber
Z_{N}&=Z_{1}^{N}\Big(\prod_{n=1}^{\infty}(1-Q_{\rho}^n)\Big)^{N-1}\,\prod_{1\leq a<b\leq N}\frac{G_{2}(Q_{ab}Q_{m}\sqrt{t\,q};\rho,\epsilon_1,-\epsilon_2)G_{2}(Q_{ab}\,Q_{m}^{-1}\sqrt{t\,q};\rho,\epsilon_1,-\epsilon_2)}
{G_{2}(Q_{ab}\,t;\rho,\epsilon_1,-\epsilon_2)G_{2}(Q_{ab}\,q;\rho,\epsilon_1,-\epsilon_2)}\\\label{pp2}
&=Z_{1}^{N}\Big(\prod_{n=1}^{\infty}(1-Q_{\rho}^n)\Big)^{N-1}\,\prod_{1\leq a<b\leq N}\frac{G_{2}(Q_{ab};\rho,\epsilon_1,\epsilon_2)G_{2}(Q_{\rho}Q_{ab}^{-1};\tau,\epsilon_1,\epsilon_2)}
{G_{2}(Q_{ab}Q_{m}\sqrt{t\,q};\rho,\epsilon_1,\epsilon_2)G_{2}(Q_{ab}\,Q_{m}^{-1}\sqrt{t\,q};\rho,\epsilon_1,\epsilon_2)}
\end{align}
with the explicit expression
\begin{align}
Z_{1}=\Big(\prod_{n=1}^{\infty}(1-Q_{\rho}^{n})^{-1}\Big)\,\prod_{i,j=1}^{\infty}\frac{(1-Q_{\rho}^{n-1}\,Q_{m}\,q^{i-\frac{1}{2}}\,t^{j-\frac{1}{2}})
(1-Q_{\rho}^{n}Q_{m}^{-1}\,q^{i-\frac{1}{2}}t^{j-\frac{1}{2}})}{
(1-Q_{\rho}^{n}q^{i}t^{j-1})(1-Q_{\rho}^{n}q^{i-1}t^{j})}\,,\label{u1pf}
\end{align}

Using Eq.(\ref{pp2}) and the modular transformation satisfied by $G_{2}(z;\rho,\epsilon_1,\epsilon_2)$ we see that:
\begin{align}
\fbox{$\displaystyle{\frac{Z_{N}(\tau,\epsilon_1,\epsilon_2)}{Z_{N}(\frac{\tau}{\epsilon_1},-\frac{1}{\epsilon_1},\frac{\epsilon_2}{\epsilon_1})Z_{N}(\frac{\tau}{\epsilon_2},\frac{\epsilon_1}{\epsilon_2},-\frac{1}{\epsilon_2})}=Z_{N}\Big(-\frac{1}{\tau},\frac{\epsilon_1}{\tau},\frac{\epsilon_2}{\tau}\Big)\,.}$}
\end{align}
Thus this class of elliptic Calabi-Yau threefold do indeed satisfy the non-perturbative modular transformation of Lockhart and Vafa \cite{Lockhart:2012vp}.

\section{Discussion}
In this paper we have shown that there is a simple class of elliptic Calabi-Yau threefolds for which the product representation of the refined topological string partition function can be determined using the relation between Schur functions and free fermionic Fock space. The product  representation allows us to study the non-perturbative modular transformations of \cite{Lockhart:2012vp}. The class of elliptic Calabi-Yau threefolds discussed in this paper were simplest in the sense that they had no compact 4-cycles. It would be interesting to see of the non-perturbative modular transformation holds for an elliptic Calabi-Yau threefold with compact 4-cycles. We hope to report on this in the future \cite{coming}.

\end{document}